\documentclass[sn-mathphys,Numbered]{sn-jnl}

\usepackage[utf8]{inputenc}
\usepackage{graphicx}
\usepackage{amsmath,amssymb,amsfonts,bm}
\usepackage{mathtools}
\usepackage{booktabs}
\usepackage{array}
\usepackage{xcolor}
\usepackage{textcomp}
\usepackage{manyfoot}
\usepackage{siunitx}
\usepackage{float}
\usepackage{subcaption}
\usepackage{geometry}
\usepackage{enumitem}
\usepackage{url}

\graphicspath{{figures/}}

\AtBeginDocument{\hypersetup{
  pdftitle={A consistent coupled level-set and volume-of-fluid solver for two-phase flow in OpenFOAM},
  pdfauthor={Amirhossein Taran}}}

\newcommand{\solver}{\texttt{interLevelSetFoam}}

\begin{document}

\title[A consistent coupled level-set/VOF solver for OpenFOAM]{A consistent coupled level-set and volume-of-fluid solver for two-phase flow in OpenFOAM}

\author*[1]{\fnm{Amirhossein} \sur{Taran}}\email{amirhossein.taran@ucd.ie}

\affil[1]{\orgdiv{School of Mechanical and Materials Engineering}, \orgname{University College Dublin}, \orgaddress{\country{Ireland}}}

\abstract{\unboldmath
Coupled level-set and volume-of-fluid methods combine the mass conservation of a transported volume fraction with the smooth interface description of a signed-distance function. In the simple coupled variant, the level set is rebuilt from the volume fraction during every time step rather than transported, and it supplies the interface curvature and the fluid properties. This paper presents an implementation of the simple coupled method in OpenFOAM and sets out four conditions that such an implementation must satisfy to stay consistent. The density in the time derivative and in the convective mass flux of the momentum equation must be the same field, the curvature in the surface tension force must come from the level set, the blended kinematic viscosity must reproduce the blended dynamic viscosity, and the smoothed fields must be evaluated on boundary and processor patches as well as in the cells. The solver is tested on the second two-dimensional rising-bubble case of Hysing et al. on two meshes. Up to $t = 1.75~\mathrm{s}$, while the three reference codes of that benchmark agree, the rise velocity differs from the reference by at most $0.0064$ and $0.0073~\mathrm{m/s}$ on the two meshes, and its first maximum is $1.3\%$ and $2.0\%$ below the reference. These differences grow slightly with refinement. The same happens when the curvature and the properties are taken from the volume fraction, which reproduces the standard OpenFOAM solver, so their cause lies outside the coupling. Over the same interval the level-set coupling stays closer to the reference than this volume-fraction coupling, with a largest velocity difference of $0.0073$ against $0.0094~\mathrm{m/s}$ on the finer mesh. The bubble volume is conserved to the printed precision, and a run on four processors matches the serial run. The solver is released under the GNU General Public License.
}


\maketitle

\section{Introduction}
\label{sec:intro}

\noindent
Bubbles, drops, and free surfaces are usually simulated on a fixed mesh with an interface-capturing method, in which a scalar field marks which fluid occupies each cell. The volume-of-fluid (VOF) method of \citet{hirt_volume_1981} transports the volume fraction of one fluid, so the mass of each phase is conserved by the discrete transport. However, the volume fraction changes sharply across the interface, and the normal and curvature computed from its gradient are noisy. In the continuum surface force model of \citet{brackbill_continuum_1992}, that curvature sets the surface tension force, so curvature errors drive spurious currents near the interface \citep{francois_balanced_2006,popinet_numerical_2018}. The level-set method \citep{osher_fronts_1988,sussman_level_1994} instead represents the interface as the zero contour of a signed-distance function. The function is smooth, so its normal and curvature are well defined, but it is not conserved under advection and reinitialisation, and the enclosed volume drifts over time.

\noindent
Coupled level-set and volume-of-fluid methods use both fields. \citet{sussman_coupled_2000} advected the level set and the volume fraction together and reinitialised the level set from the interface reconstructed from the volume fraction. \citet{menard_coupling_2007} combined a similar coupling with a ghost-fluid treatment of the pressure jump and applied it to the primary break-up of a liquid jet. \citet{albadawi_influence_2013} proposed a simpler coupling for the OpenFOAM VOF solver, in which only the volume fraction is transported and the level set is rebuilt from it at every time step. They used it to study how the implementation of surface tension influences the predicted growth and detachment of bubbles. \citet{yamamoto_validation_2017} later combined this simple coupled method with a density-scaled balanced surface force model and validated it for multiphase systems with thermocapillary flows.

\noindent
The simple coupled method is attractive because it leaves the volume-fraction transport untouched and changes only how the curvature and the fluid properties are computed. It also introduces a second description of the same interface, and every quantity that enters the momentum equation must then be taken consistently from one of the two. An implementation can conserve mass in the transport step and still assemble a momentum equation whose terms see two different interfaces. This paper sets out four consistency conditions for the simple coupled method and implements them in an OpenFOAM solver, \solver{}, built on the \texttt{interFoam} VOF solver \citep{weller_tensorial_1998,deshpande_evaluating_2012}. The solver is assessed on the two-dimensional rising-bubble benchmark of \citet{hysing_quantitative_2009}, against the reference solutions of that benchmark on two meshes, against the volume-fraction coupling on the same meshes, and in parallel.

\section{Method}
\label{sec:method}

\subsection{Governing equations}
\label{sec:governing}

\noindent
Two incompressible, isothermal, immiscible fluids are described as a single fluid with variable properties. The volume fraction $\alpha$ of fluid 1 is transported by
\begin{equation}
\frac{\partial \alpha}{\partial t} + \nabla \cdot (\mathbf{U} \alpha) + \nabla \cdot \left[ \mathbf{U}_r \alpha (1 - \alpha) \right] = 0,
\label{eq:alpha}
\end{equation}
where $\mathbf{U}$ is the mixture velocity and $\mathbf{U}_r = c_\alpha |\mathbf{U}| \mathbf{n}$ is an artificial compression velocity normal to the interface, scaled by the coefficient $c_\alpha$. Equation~\eqref{eq:alpha} is solved with the MULES limiter of \texttt{interFoam}, which keeps $\alpha$ between 0 and 1 \citep{deshpande_evaluating_2012}. The velocity satisfies $\nabla \cdot \mathbf{U} = 0$ and the momentum equation
\begin{equation}
\frac{\partial (\rho \mathbf{U})}{\partial t} + \nabla \cdot (\dot{\mathbf{m}} \mathbf{U}) - \mathbf{U} \left( \frac{\partial \rho}{\partial t} + \nabla \cdot \dot{\mathbf{m}} \right) = - \nabla p_{rgh} - (\mathbf{g} \cdot \mathbf{x}) \nabla \rho + \nabla \cdot \left[ \mu \left( \nabla \mathbf{U} + \nabla \mathbf{U}^{T} \right) \right] + \mathbf{f}_\sigma,
\label{eq:momentum}
\end{equation}
where $\rho$ is the density, $\dot{\mathbf{m}}$ is the mass flux, $\mu$ is the dynamic viscosity, $p_{rgh} = p - \rho\, \mathbf{g} \cdot \mathbf{x}$ is the pressure without its hydrostatic part, $\mathbf{g}$ is the gravitational acceleration, and $\mathbf{f}_\sigma$ is the surface tension force per unit volume. The bracketed term on the left is zero when $\rho$ and $\dot{\mathbf{m}}$ satisfy the continuity equation of the mixture. Its discrete value is the mass imbalance of the two fields. Subtracting it removes the spurious source $\mathbf{U}(\partial \rho/\partial t + \nabla \cdot \dot{\mathbf{m}})$ from the conservative form, so that for $\dot{\mathbf{m}} \approx \rho \mathbf{U}$ the left-hand side reduces to $\rho\, \mathrm{D}\mathbf{U}/\mathrm{D}t$ whatever the size of the imbalance. The term is new relative to \texttt{interFoam}, and the reason for it is given in Section~\ref{sec:consistency}.

\subsection{Rebuilding the level set}
\label{sec:levelset}

\noindent
Following \citet{albadawi_influence_2013}, the level set $\phi$ is rebuilt from the current volume fraction at every outer corrector of the time step. The lengths that control the rebuild are set as multiples of a reference cell size $h$. First, an initial distance is assigned from the volume fraction as
\begin{equation}
\phi_0 = (2 \alpha - 1)\, \Gamma, \qquad \Gamma = 0.75\, h,
\label{eq:seed}
\end{equation}
which is positive in fluid 1 and negative in fluid 2. Second, $\phi_0$ is moved towards a signed distance by the reinitialisation equation of \citet{sussman_level_1994},
\begin{equation}
\frac{\partial \phi}{\partial \tau} = S(\phi_0) \left( 1 - |\nabla \phi| \right), \qquad \phi(\tau = 0) = \phi_0,
\label{eq:redistance}
\end{equation}
in the pseudo-time $\tau$, with $S(\phi_0) = \phi_0 / |\phi_0|$ and $S(0) = 0$. The interface is given a half-width $\varepsilon = 1.5\, h$. Equation~\eqref{eq:redistance} is advanced with explicit Euler steps of $\Delta \tau = 0.1\, h$, with $\nabla \phi$ evaluated by the central Gauss linear scheme, for $N_\tau = \lceil \varepsilon / \Delta \tau \rceil = 15$ steps. The distance is therefore corrected over a band of half-width $\varepsilon$ around the interface. The central gradient does not make $\phi$ an exact distance, and the solver reports the mean of $|1 - |\nabla \phi||$ in the band at every rebuild. In the runs of Section~\ref{sec:results} it is between 0.1 and 0.2. Third, the smoothed Heaviside function and the interface curvature are evaluated from $\phi$ as
\begin{equation}
H(\phi) =
\begin{cases}
0, & \phi < -\varepsilon, \\
\dfrac{1}{2} \left[ 1 + \dfrac{\phi}{\varepsilon} + \dfrac{1}{\pi} \sin \left( \dfrac{\pi \phi}{\varepsilon} \right) \right], & |\phi| \le \varepsilon, \\
1, & \phi > \varepsilon,
\end{cases}
\label{eq:heaviside}
\end{equation}
and
\begin{equation}
\kappa = - \nabla \cdot \left( \frac{\nabla \phi}{|\nabla \phi|} \right).
\label{eq:curvature}
\end{equation}
The curvature is computed from the face-interpolated gradient of $\phi$, normalised on each face and summed over the faces of each cell. The surface tension force is
\begin{equation}
\mathbf{f}_\sigma = \sigma \kappa \nabla H = \sigma \kappa\, \delta(\phi) \nabla \phi,
\label{eq:surfacetension}
\end{equation}
where $\sigma$ is the surface tension coefficient and $\delta = \mathrm{d}H/\mathrm{d}\phi$ is the smoothed Dirac delta. As in \texttt{interFoam}, Eq.~\eqref{eq:surfacetension} is evaluated on the cell faces, as $\sigma \kappa_f$ times the face-normal gradient of $H$, so that it enters the pressure equation in the same discrete form as the pressure gradient \citep{francois_balanced_2006}. The density and the dynamic viscosity are blended with the same Heaviside function,
\begin{equation}
\rho = \rho_1 H + \rho_2 (1 - H), \qquad \mu = \mu_1 H + \mu_2 (1 - H),
\label{eq:properties}
\end{equation}
where the subscripts 1 and 2 denote the two fluids.

\subsection{Consistency conditions}
\label{sec:consistency}

\noindent
The volume fraction and the level set describe the same interface, but they are not identical, because $H$ changes smoothly over a band of width $2\varepsilon$ around the interface while $\alpha$ is transported as a separate field. Each term of Eq.~\eqref{eq:momentum} must therefore take its interface from one of the two fields consistently, which leads to the conditions (C1) to (C4) below.

\begin{enumerate}[label=(C\arabic*), leftmargin=*]

\item \textbf{The density in the time derivative and in the mass flux must be the same field.} In \texttt{interFoam}, the face mass flux is built inside the MULES step from the limited volume-fraction flux, so that it satisfies the discrete continuity equation together with $\rho = \rho_1 \alpha + \rho_2 (1 - \alpha)$. Once $\rho$ is taken from Eq.~\eqref{eq:properties}, that mass flux no longer matches it, and $\partial(\rho \mathbf{U})/\partial t$ and $\nabla \cdot (\dot{\mathbf{m}} \mathbf{U})$ see two different interfaces. The present solver builds the face mass flux from the same Heaviside function, $\dot{m}_f = \left[ \rho_2 + (\rho_1 - \rho_2) H_f \right] F_f$, where $F_f$ is the volumetric face flux and $H_f$ is the face-interpolated Heaviside function. Because $H$ is rebuilt rather than transported, the pair still does not satisfy the continuity equation exactly, and the remaining imbalance is removed by the bracketed term of Eq.~\eqref{eq:momentum}.

\item \textbf{The curvature in the surface tension force must be the level-set curvature.} The VOF mixture model in OpenFOAM computes its own curvature from $\alpha$. If the level set is rebuilt but the force is still taken from that model, the surface tension is the same as in the VOF method and the level set has no effect on it. The same force must also be used in the momentum predictor and in the pressure equation.

\item \textbf{The blended kinematic viscosity must reproduce the blended dynamic viscosity.} The OpenFOAM viscous stress models take the kinematic viscosity $\nu$ and form $\mu = \rho \nu$. Blending $\nu$ linearly with $H$ while $\rho$ is also blended does not give the $\mu$ of Eq.~\eqref{eq:properties}. For the fluids of Section~\ref{sec:setup} at $H = 1/2$, the product of the two linear blends is $27.5~\mathrm{Pa\,s}$, whereas the blended dynamic viscosity is $5.05~\mathrm{Pa\,s}$. The present solver sets $\nu = \mu / \rho$ with both taken from Eq.~\eqref{eq:properties}.

\item \textbf{The smoothed fields must be evaluated on boundary and processor patches.} Face interpolation of $\rho$ and the face-normal gradient of $H$ use the patch values at walls and at the boundaries between subdomains of a parallel run. If $H$ is computed only in the cells and its patch values keep their initial value, a wall in contact with fluid 1 carries the density of fluid 2. The hydrostatic term $(\mathbf{g} \cdot \mathbf{x}) \nabla \rho$ then acts on the wall faces as if a density jump sat on the wall, and the same error at processor boundaries makes the solution depend on the domain decomposition.

\end{enumerate}

\subsection{Implementation}
\label{sec:implementation}

\noindent
The solver \solver{} is built on \texttt{interFoam} and has been compiled with OpenFOAM v2306, v2312, v2512, and v2606. The level-set steps are contained in one class that extends the immiscible two-phase mixture model of OpenFOAM, so the volume-fraction transport, the turbulence models, and the dynamic-mesh support of \texttt{interFoam} are used unchanged. Within each outer corrector of the PIMPLE loop, the volume fraction is first advanced by Eq.~\eqref{eq:alpha}, including all of its MULES corrections. The level set is then rebuilt once from the new $\alpha$, after which the Heaviside function, the curvature, and the viscosity are updated and $\rho$ and $\dot{m}_f$ are set from $H$. The momentum and pressure equations follow as in \texttt{interFoam}, with the surface tension force of Eq.~\eqref{eq:surfacetension}. The Heaviside function is evaluated as $\frac{1}{2}[1 + x + \sin(\pi x)/\pi]$ with $x = \phi/\varepsilon$ limited to $[-1, 1]$. This expression equals Eq.~\eqref{eq:heaviside} everywhere and is applied to cells and patches alike. All parameters are read from an optional dictionary with the defaults given above, and $h$ defaults to the smallest distance between neighbouring cell centres. The curvature and the fluid properties can each be switched back to the volume fraction, which gives the volume-fraction coupling used for comparison in Section~\ref{sec:results}.

\section{Results}
\label{sec:results}

\subsection{Rising-bubble benchmark}
\label{sec:setup}

\noindent
The test case is the second rising-bubble case of \citet{hysing_quantitative_2009}, chosen because its density ratio of 1000 makes it sensitive to how the density enters the momentum equation. A circular bubble of fluid 2 with a radius of $0.25~\mathrm{m}$ starts at rest at $(0.5, 0.5)~\mathrm{m}$ in a closed $1 \times 2~\mathrm{m}$ box filled with fluid 1. The initial volume fraction is set with the \texttt{setAlphaField} utility of OpenFOAM, which computes the fraction of each cell cut by the circle. Table~\ref{tab:setup} lists the fluid properties, which give a Reynolds number of 35 and an E\"otv\"os number of 125. The top and bottom walls are no-slip and the side walls are free-slip. The pressure $p_{rgh}$ has a fixed-flux condition on all walls, so that no fluid crosses them. Two uniform meshes are used, with $1/h = 80$ and $1/h = 160$ cells per metre, and the time step is fixed at the values in Table~\ref{tab:setup}. The time derivatives use the Euler scheme, the momentum convection uses linear-upwind interpolation, and $\alpha$ is convected with the van Leer limiter and the compression coefficient $c_\alpha = 1$. Each time step has one PIMPLE outer corrector with three pressure correctors and no momentum predictor, and the MULES step uses two volume-fraction correctors. The level-set parameters are those of Section~\ref{sec:levelset}. All runs use OpenFOAM v2606. The $1/h = 160$ runs use four processors, and the $1/h = 80$ runs are serial unless stated otherwise.

\begin{table}[tbp]
\centering
\small
\caption{Parameters of the rising-bubble benchmark, test case 2 of \citet{hysing_quantitative_2009}. Fluid 1 surrounds the bubble and fluid 2 fills it.}
\label{tab:setup}
\begin{tabular}{l c c}
\toprule
Quantity & Symbol & Value \\
\midrule
Density of fluid 1, fluid 2 & $\rho_1$, $\rho_2$ & $1000$, $1~\mathrm{kg/m^3}$ \\
Dynamic viscosity of fluid 1, fluid 2 & $\mu_1$, $\mu_2$ & $10$, $0.1~\mathrm{Pa\,s}$ \\
Surface tension coefficient & $\sigma$ & $1.96~\mathrm{N/m}$ \\
Gravitational acceleration & $|\mathbf{g}|$ & $0.98~\mathrm{m/s^2}$ \\
Initial bubble radius & $r_0$ & $0.25~\mathrm{m}$ \\
Mesh spacing & $h$ & $1/80$, $1/160~\mathrm{m}$ \\
Time step & $\Delta t$ & $7.8125 \times 10^{-4}$, $3.90625 \times 10^{-4}~\mathrm{s}$ \\
\bottomrule
\end{tabular}
\end{table}

\noindent
The centre of mass $y_c$ and the rise velocity $V_c$ of the bubble are defined by \citet{hysing_quantitative_2009} as averages of the vertical position and the vertical velocity over the region occupied by fluid 2. In the solver they are evaluated at every time step as averages weighted by $1 - \alpha$. The first maximum of $V_c$ is the largest value before $t = 1.3~\mathrm{s}$ and the second is the largest value between $t = 1.5$ and $2.6~\mathrm{s}$. \citet{hysing_quantitative_2009} solved this case with three codes, TP2D, FreeLIFE, and MooNMD, and their solutions are available from the benchmark website \citep{featflow_bubble}. The TP2D solution on its finest mesh, $1/h = 640$, is used as the reference, and the spread of the three codes on their finest meshes indicates how well the reference itself is known. Hysing et al.\ reported that the three codes agree closely up to $t = 1.75$ to $2.0~\mathrm{s}$, after which the trailing skirts of the bubble break up and the codes predict different shapes. They also described the second maximum of $V_c$ as more ambiguous than the first, and the final centre of mass as of limited meaning because the final shape of the bubble is unknown. The comparison below therefore concentrates on the time up to $t = 1.75~\mathrm{s}$ and on the first maximum. Table~\ref{tab:results} collects the benchmark quantities of all runs.

\begin{table}[tbp]
\centering
\small
\caption{Benchmark quantities for test case 2. $V_{c,1}$ and $V_{c,2}$ are the first and second maxima of the rise velocity, with the time of each in seconds in brackets, and $y_c$ is the centre of mass at $t = 3~\mathrm{s}$. The reference rows are from Tables XIII and XVI of \citet{hysing_quantitative_2009}. MooNMD uses a moving interface-fitted mesh, so no mesh spacing is given for it. Differences quoted in the text are computed from the unrounded data.}
\label{tab:results}
\setlength{\tabcolsep}{5pt}
\begin{tabular}{l c c c c}
\toprule
Run & $1/h$ & $V_{c,1}$ (m/s) & $V_{c,2}$ (m/s) & $y_c$ (m) \\
\midrule
TP2D & $80$ & $0.2638$ ($0.725$) & $0.2597$ ($1.969$) & $1.1370$ \\
TP2D & $160$ & $0.2570$ ($0.743$) & $0.2522$ ($2.023$) & $1.1377$ \\
TP2D (reference) & $640$ & $0.2524$ ($0.733$) & $0.2434$ ($2.071$) & $1.1380$ \\
FreeLIFE & $160$ & $0.2514$ ($0.728$) & $0.2440$ ($1.984$) & $1.1249$ \\
MooNMD & & $0.2502$ ($0.732$) & $0.2393$ ($2.060$) & $1.1376$ \\
\midrule
Present & $80$ & $0.2491$ ($0.734$) & $0.2287$ ($1.913$) & $1.1140$ \\
Present, no imbalance term & $80$ & $0.2497$ ($0.714$) & $0.2295$ ($1.886$) & $1.1150$ \\
Volume-fraction coupling & $80$ & $0.2488$ ($0.730$) & $0.2278$ ($1.923$) & $1.1100$ \\
\texttt{interFoam} & $80$ & $0.2488$ ($0.730$) & $0.2278$ ($1.923$) & $1.1100$ \\
\midrule
Present & $160$ & $0.2472$ ($0.738$) & $0.2315$ ($1.982$) & $1.1174$ \\
Volume-fraction coupling & $160$ & $0.2471$ ($0.734$) & $0.2309$ ($2.001$) & $1.1140$ \\
\bottomrule
\end{tabular}
\end{table}

\subsection{Comparison with the reference solution}
\label{sec:reference}

\begin{figure}[tbp]
\centering
\includegraphics[width=\textwidth]{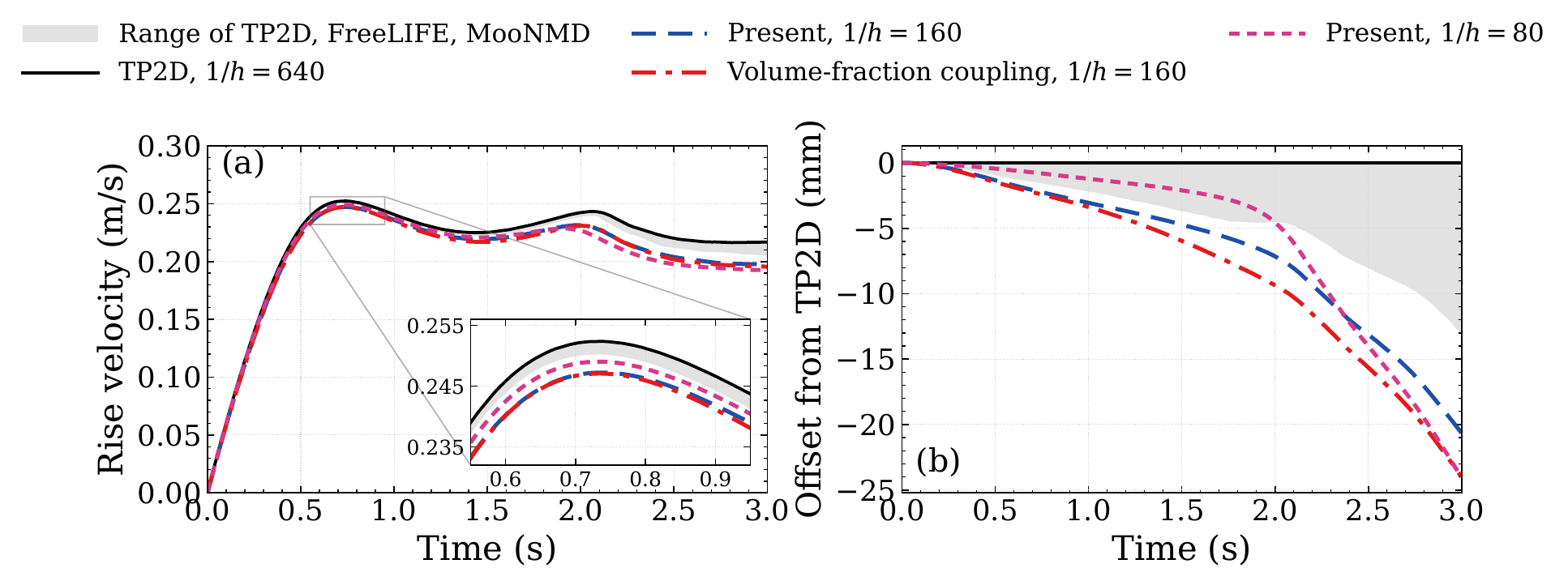}
\caption{Comparison with the reference solutions of \citet{hysing_quantitative_2009}. The grey band is the range of the three benchmark codes on their finest meshes. (a) Rise velocity, with the first maximum enlarged in the inset. (b) Offset of the centre of mass from the TP2D reference.}
\label{fig:benchmark}
\end{figure}

\noindent
Figure~\ref{fig:benchmark}a compares the rise velocity of the present solver with the reference. Up to $t = 1.75~\mathrm{s}$, where the three reference codes agree, the present rise velocity differs from TP2D by at most $0.0064~\mathrm{m/s}$ on the $1/h = 80$ mesh and $0.0073~\mathrm{m/s}$ on the $1/h = 160$ mesh. The first maximum is $0.2491~\mathrm{m/s}$ at $t = 0.734~\mathrm{s}$ and $0.2472~\mathrm{m/s}$ at $t = 0.738~\mathrm{s}$ on the two meshes, against $0.2524~\mathrm{m/s}$ at $t = 0.733~\mathrm{s}$ for the reference. Its time agrees within $0.005~\mathrm{s}$, and its value is $1.3\%$ and $2.0\%$ low. On the finer mesh it lies $1.2\%$ below the lowest of the three codes, MooNMD.

\noindent
Refining the mesh moves the solution slightly away from the reference in this interval, although the TP2D values on the same two meshes approach it. From $1/h = 80$ to $160$, the shortfall at the first maximum grows from $1.3\%$ to $2.0\%$, the largest velocity difference grows from $0.0064$ to $0.0073~\mathrm{m/s}$, and the offset of the centre of mass at $t = 1.75~\mathrm{s}$ grows from $2.8$ to $5.7~\mathrm{mm}$ (Fig.~\ref{fig:benchmark}b). The volume-fraction coupling behaves in the same way, with a first maximum within $0.0003~\mathrm{m/s}$ of the level-set coupling on both meshes. The cause therefore lies in a part of the solver that both couplings share, and the time step, which is refined together with the mesh, is one candidate that this study did not vary.

\noindent
After $t = 1.75~\mathrm{s}$ the differences grow. Up to $t = 2.0~\mathrm{s}$, the upper end of the interval in which Hysing et al.\ found agreement, the largest velocity difference is $0.0109~\mathrm{m/s}$ on the $1/h = 160$ mesh and $0.0158~\mathrm{m/s}$ on the $1/h = 80$ mesh. The second maximum is $0.2315~\mathrm{m/s}$ at $t = 1.982~\mathrm{s}$ on the finer mesh, $4.9\%$ below the reference and $3.3\%$ below the lowest of the three codes. At $t = 3~\mathrm{s}$ the centre of mass on the finer mesh is $1.1174~\mathrm{m}$, $20.7~\mathrm{mm}$ below TP2D and below all three codes. Relative to the distance the bubble has risen, the offset from TP2D is $3.2\%$.

\begin{figure}[tbp]
\centering
\includegraphics[width=0.58\textwidth]{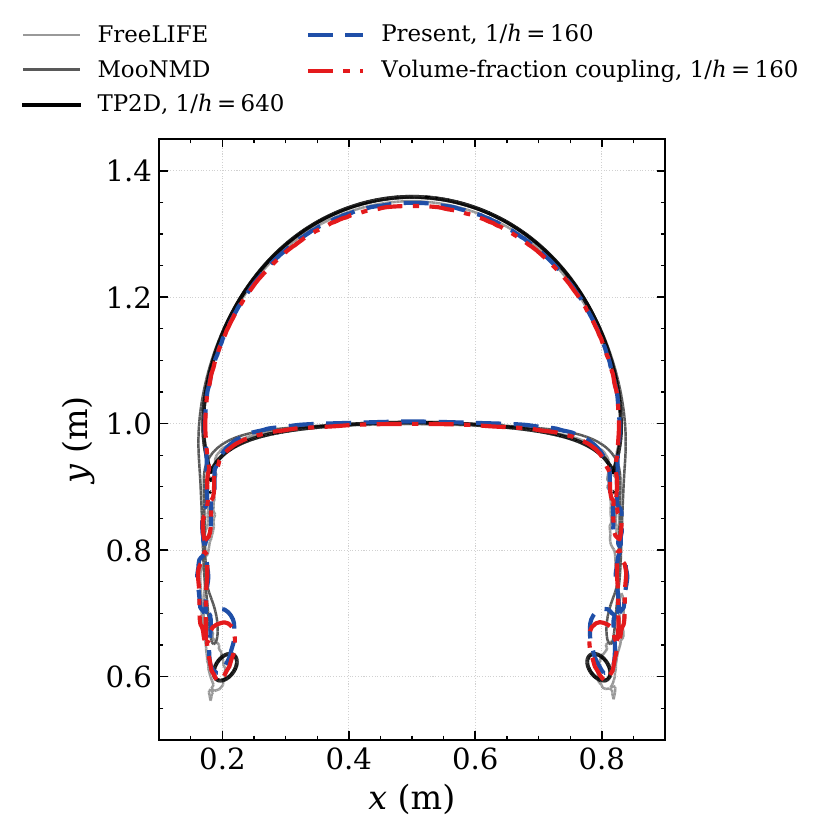}
\caption{Bubble shape at $t = 3~\mathrm{s}$. The present solver and the volume-fraction coupling are shown as the $\alpha = 0.5$ contour on the $1/h = 160$ mesh. The three benchmark codes are shown on their finest meshes \citep{hysing_quantitative_2009}.}
\label{fig:shape}
\end{figure}

\noindent
Figure~\ref{fig:shape} compares the bubble shapes at $t = 3~\mathrm{s}$. On the vertical centreline the cap of the present solution is at $y = 1.349~\mathrm{m}$, against $1.352$ to $1.359~\mathrm{m}$ for the three codes, and its lower surface is at $1.003~\mathrm{m}$, against $0.999$ to $1.002~\mathrm{m}$. The trailing skirts differ more. In TP2D they have broken off into satellite droplets near $y = 0.6~\mathrm{m}$. On the $1/h = 160$ mesh the skirts of the present solution have also broken off, on each side into a filament one cell wide between $y = 0.70$ and $0.79~\mathrm{m}$ and a droplet between $y = 0.61$ and $0.70~\mathrm{m}$, and the volume-fraction coupling gives the same picture. On the $1/h = 80$ mesh the skirts are still attached to the bubble at $t = 3~\mathrm{s}$. FreeLIFE and MooNMD give two further shapes for the skirts, which illustrates the remark of \citet{hysing_quantitative_2009} that the codes differ after break-up.

\subsection{Coupling options and parallel run}
\label{sec:options}

\noindent
With both the curvature and the properties taken from the volume fraction, the solver reproduces \texttt{interFoam} on the $1/h = 80$ mesh to all digits in Table~\ref{tab:results}, which confirms that the added mass-imbalance term has no visible effect when $\rho$ and $\dot{m}_f$ both come from the MULES step. Up to $t = 1.75~\mathrm{s}$, the level-set coupling stays closer to the reference than this volume-fraction coupling on both meshes. Its largest velocity difference from TP2D is $0.0073~\mathrm{m/s}$ on the $1/h = 160$ mesh, against $0.0094~\mathrm{m/s}$, and its centre of mass at $t = 1.75~\mathrm{s}$ is $5.7~\mathrm{mm}$ below TP2D, against $7.6~\mathrm{mm}$. On the $1/h = 80$ mesh the corresponding values are $0.0064$ and $0.0082~\mathrm{m/s}$, and $2.8$ and $4.4~\mathrm{mm}$. The difference persists to $t = 3~\mathrm{s}$, where the centre of mass on the finer mesh is $20.7~\mathrm{mm}$ below TP2D for the level-set coupling and $24.0~\mathrm{mm}$ for the volume-fraction coupling. The mass-imbalance term was also removed from the level-set coupling, in a separate build of the solver, on the $1/h = 80$ mesh. This moved the solution slightly closer to the reference, by $0.5~\mathrm{mm}$ in the centre of mass at $t = 1.75~\mathrm{s}$ and $1.1~\mathrm{mm}$ at $t = 3~\mathrm{s}$, so in this case the term gives no measurable benefit. It is kept because it makes the discrete momentum equation consistent with the mass flux whatever the size of the imbalance. In every run the volume fraction of fluid 1 printed by the solver to six significant digits is the same at $t = 0$ and $t = 3~\mathrm{s}$, so the bubble volume changes by less than $0.001\%$.

\noindent
The $1/h = 80$ case was also run on four processors with a $2 \times 2$ decomposition. Over the whole run, the centre of mass differs from the serial result by at most $2.9 \times 10^{-6}~\mathrm{m}$ and the rise velocity by at most $1.8 \times 10^{-5}~\mathrm{m/s}$, which confirms that the smoothed fields are consistent across processor boundaries.

\section{Conclusions}
\label{sec:conclusion}

\noindent
A simple coupled level-set and volume-of-fluid solver was implemented in OpenFOAM, in which the volume fraction is transported by the MULES scheme of \texttt{interFoam} and a signed-distance function is rebuilt from it at every outer corrector to supply the curvature, the density, and the viscosity. Four conditions were set out that keep the two descriptions of the interface consistent in the momentum equation. The density in the time derivative and in the mass flux must be the same field, the surface tension force must use the level-set curvature, the blended kinematic viscosity must reproduce the blended dynamic viscosity, and the smoothed fields must be evaluated on boundary and processor patches. A mass-imbalance term was added to the momentum equation, so that it stays consistent although the rebuilt density is not transported exactly by the mass flux.

\noindent
The solver was tested on the second rising-bubble case of \citet{hysing_quantitative_2009} on meshes with $1/h = 80$ and $160$. Up to $t = 1.75~\mathrm{s}$, where the three reference codes agree, the rise velocity differed from the TP2D reference by at most $0.0064$ and $0.0073~\mathrm{m/s}$ on the two meshes. The first maximum of the rise velocity was $1.3\%$ and $2.0\%$ below the reference, at the same time within $0.005~\mathrm{s}$. These differences grew slightly with refinement, and the volume-fraction coupling, which reproduced \texttt{interFoam}, showed the same trend, so they do not come from the level-set coupling. In the same interval the level-set coupling stayed closer to the reference than the volume-fraction coupling, with largest velocity differences of $0.0073$ against $0.0094~\mathrm{m/s}$ on the finer mesh. At $t = 3~\mathrm{s}$ the cap and the lower surface of the bubble were within $10~\mathrm{mm}$ of the three reference codes, and on the finer mesh the trailing skirts had broken off, as in TP2D. The bubble volume was conserved to the printed precision in every run, and a run on four processors matched the serial run to $2.9 \times 10^{-6}~\mathrm{m}$ in the centre of mass.

\noindent
The runs test a solver that satisfies the four conditions. Apart from the mass-imbalance term, they do not isolate the effect of violating each condition. Removing that term changed the centre of mass by about $1~\mathrm{mm}$ and gave no measurable loss of accuracy in this case. The source of the small shortfall that grows with refinement remains to be identified, and the time step, which was refined together with the mesh, is the first candidate to test.


\backmatter

\bmhead{Code availability}
\noindent
The solver and the rising-bubble tutorial are available at \url{https://github.com/cloner0110/LevelSet-VOF} under the GNU General Public License v3.

\bmhead{Conflict of interest statement}
\noindent
The author has no conflicts to disclose.

\bibliography{bibliography}

\end{document}